\documentclass[twocolumn,aps,prd,nofootinbib,floatfix, preprintnumbers]{revtex4}
\usepackage{graphicx}
\usepackage{amsmath}
\usepackage{amssymb}
\usepackage{bm}

\newcommand{\bs}[1]{\boldsymbol{{#1}}}

\begin{document}

\title{Unravelling  $t\bar t h$ via the matrix element  method} 

\preprint{NIKHEF-2013-011, CP3-13-14}

\author{Pierre Artoisenet$^{a}$, Priscila de Aquino$^{b}$, Fabio Maltoni$^{c}$, Olivier Mattelaer$^{c,d}$}

\affiliation{
$^{a}$ Nikhef Theory Group, Science Park 105, 1098 XG Amsterdam, The Netherlands\\
$^{b}$Theoretische Natuurkunde and IIHE/ELEM, 
 Vrije Universiteit Brussel, and International Solvay Institutes, Pleinlaan 2,  B-1050 Brussels, Belgium\\
$^{c}$Centre for Cosmology, Particle Physics and Phenomenology (CP3), Universit\'e catholique de Louvain, Chemin du Cyclotron 2, B-1348 Louvain-la-Neuve, Belgium\\
$^{d}$Department of Physics, University of Illinois at Urbana-Champaign,  1110 West Green Street, Urbana, IL  61801 }

\date{\today}
\begin{abstract}
Associated production of the Higgs boson with a top-antitop pair is a key channel to gather further information on  the nature of the newly discovered boson at the LHC. Experimentally, however, its observation is very challenging due to the combination of  small rates, difficult multi-jet  final states and overwhelming backgrounds.  In the Standard Model the largest number of events is expected when $h\to b\bar b$, giving rise to a  $W^+W^-b\bar bb\bar b$ signature, deluged in $t\bar t$+jets. A promising strategy to improve the sensitivity  is to maximally exploit the theoretical information on the signal and background processes by means of the matrix element method.  We show how, despite the complexity of the final state, the method can be efficiently applied to discriminate the signal against combinatorial and $t\bar t$+jets backgrounds.  Remarkably, we find that  a moderate integrated luminosity in the next LHC run will be enough to make the signature involving both $W$'s decaying leptonically  as sensitive as the single-lepton one.  \end{abstract}

\maketitle

\section{Introduction}

Evidence for the recently discovered new heavy boson to be the long-sought-for
Higgs particle of the Standard Model is already quite compelling~\cite{cms2013,atlas2013}. 
Rates and distributions are compatible with the predictions of a scalar particle coupling to other
SM particles with a strength proportional to their mass. The current sensitivities
and accuracies of the golden production-decay modes, however,  are not sufficient to draw a final
answer on the strength and the structure of the couplings without additional 
hypotheses. Other channels need to be investigated. 

Prominent among the yet-to-be-explored production modes is the $t\bar t h$ 
associated production. The main interest of this channel stems from the fact that the rate is manifestly 
proportional to the square of the SM Yukawa coupling to the top quarks. In addition,
more differential observables could bring information on the coupling structure~\cite{Degrande:2012gr} and 
on the Higgs parity~\cite{Frederix:2011zi}. This channel, however, is notoriously
difficult for several reasons. The first is that production rates at hadron colliders 
are quite small due to the need of a  large cms collision energy for the initial partons, strongly suppressed by parton distribution functions. Next-to-leading order calculations~\cite{Beenakker:2001rj,Dawson:2002tg,Frederix:2011zi,Garzelli:2011vp} predict a SM rate of $0.137$ pb and $0.632$ pb with $O(10\%)$ uncertainty at the LHC for $\sqrt{s}=8$ and 14 TeV, respectively.  Current searches mainly focus on the dominant decay mode $h\to b\bar b$ and therefore on a $W^+W^-b\bar bb\bar b$ final state, other decays, such as $h\to W^+ W^-$~\cite{Maltoni:2002jr}, $h\to \tau^+ \tau^-$~\cite{Belyaev:2002ua} 
and eventually, $h\to \gamma \gamma$~\cite{Buttar:2006zd}, being much rarer demand larger integrated luminosities.
The second reason is that the  $W^+W^-b\bar bb\bar b$ signature is affected
by two different types of challenging backgrounds. On the one hand $t \bar t$ + light- or heavy-flavor jets because of the enormous rates, and on the other hand the intrinsic combinatorial background that stems from the difficulty of correctly identifying out of four $b$-jets
the two  from the Higgs decay. Finally, the complexity of the final state makes its kinematic reconstruction not straightforward mainly due to finite jet energy resolution, missing energy and the ubiquity of extra QCD radiation. 

Due to the above intrinsic difficulties, the prospects of first using this channel for discovery or just for observation have been constantly deteriorating as more accurate predictions and simulations were  available to the LHC community. More recently, the attention on this channel was revived by  Plehn et al.~\cite{Plehn:2009rk} who suggested that while drastically lowering the rates, boosted tops and Higgs in the final state would make the combinatorial background much less severe, improving the significance $S/\sqrt{B}$ of the SM Higgs observation at large enough luminosities. 

In this work we argue that the sensitivity to $t \bar t h$ can be also enhanced at low $p_T$, {\it i.e.}, where the bulk of the cross section resides, by means of the matrix element reweighting method,  improving the prospects for observation of this channel at the LHC in the coming years. 
The matrix element method is able to  efficiently reduce the combinatorial problem  for the single-lepton final states and even more for
the di-lepton final state, bringing the two to a comparable level of sensitivity already for moderate integrated luminosities.

\section{The matrix element method}

The matrix element reweigthing method (MEM), originally introduced in Ref.~\cite{Kondo:1988yd},
assigns probabilities to competing hypotheses, {\it e.g.}, signal vs. signal+background, given 
a sample of experimental events. The most attracting feature of this method is that  it makes 
maximal use of both experimental information and the theoretical model, associating a weight 
to each event based on the value of  the matrix element ({\it i.e.},  the scattering amplitude) for that specific final state configuration 
 for each of  the hypotheses.  While very simple  in its essence, in practice several technical 
and conceptual challenges arise and different level of simplifications are commonly employed.
The method, implemented using matrix elements calculated at the leading order, has been  successfully  
applied to a number of key results
in collider physics: from the most precise top mass determination~\cite{Abazov:2004cs,Abulencia:2006ry}, single top observation~\cite{Aaltonen:2009jj,Abazov:2009ii}  at the Tevatron, to the Higgs  boson discovery and characterization  at the LHC~\cite{Aad:2012tfa,Chatrchyan:2012ufa}. Efforts to include next-to-leading QCD corrections, at least for simple final states, have started~\cite{Campbell:2012cz,Campbell:2013hz}.

In this work, we define the weight associated with  an experimental event $\bs x$ given a set of hypotheses $\bs \alpha$  as
\begin{equation}
\label{weight_def}
P(\bs x |\bs \alpha)=\frac{1}{\sigma_{ \alpha}} \int d \Phi(\bs y) |M_{ \alpha}|^2 (\bs y)  W(\bs x,\bs y)\,,
\end{equation} 
where $|M_\alpha|^2(\bs y)$ is  the leading-order matrix element (giving the parton-level probability), $d \Phi(\bs y)$ is  the phase-space measure,  (including the parton distribution functions $f_1(q_1) dq_1$ and $f_2(q_2) dq_2$) and $W(\bs x, \bs y)$ is the transfer function which describes  the evolution of (the final state) parton-level configuration in $\bs y$ into a reconstructed event $\bs x$ in the detector.
The normalization by the total cross section $\sigma_{ \alpha}$  in Eq.~(\ref{weight_def}) ensures that $P(\bs x |\bs \alpha)$ is a probability density, $ \int P(\bs x | \bs \alpha) d\bs x=1$, if the transfer function is normalized to one. 

As evident from the definition in Eq.~(\ref{weight_def}), the calculation of each weight involves a non trivial multi-dimensional integration 
of complicated functions over the phase space. The problem of computing the weights for arbitrary models and processes 
was tackled in Ref.~\cite{Artoisenet:2010cn} by implementing a general algorithm in a specifically designed code named {\sc MadWeight}.
We stress that very fact of automatically, reliably and quickly calculating weights for challenging final states as those involved in $t \bar t h$  has never been achieved before. It is a significant technical result on its own that provides key evidence on the generality and flexibility of the {\sc MadWeight} approach.

One of the main limitations of the method is that the matrix elements are considered at the leading order only and therefore extra QCD radiation effects must be handled in some effective way. In our study we are inclusive on extra transverse radiation and consider only the hardest jets to be matched with the corresponding partons in the matrix element.  The transverse momentum of these partons (including isolated leptons)
is assumed to be balanced against the transverse momentum of extra radiation in the event~\cite{Alwall:2010cq}.

\section{Technical aspects}

Parton-level events for signal and backgrounds are generated with {\sc MadGraph 5}~\cite{Alwall:2011uj}, passed to {\sc Pythia 6}~\cite{Sjostrand:2006za} for showering and hadronization, employing the MLM-$k_T$  merging procedure~\cite{Alwall:2008qv}. 
Detector response simulation is  performed using {\sc Delphes 2}~\cite{Ovyn:2009tx},  with the input parameters tuned to the 
values associated with the CMS detector. Pile-up effects have not been considered.

Only the dominant background $t \bar t$+jets is taken into account, and is modeled
by generating 
parton-level $t \bar t$ processes with up to two extra partons in 
the 5-flavor scheme.
For the signal, the parton-level processes  $t \bar t h$ and $t \bar t  h + 1$ parton are considered. 
Inclusive samples for the signal and the background have been normalized to the 
total cross section at  NLO from~\cite{Frederix:2011zi} and ~\cite{Cacciari:2011hy}, respectively.
Spin correlation effects in the decays of the tops, which for signal shapes are more important than 
NLO QCD corrections~\cite{Artoisenet:2012st}, have been retained.


\begin{table}[t]
\begin{center}
\begin{tabular}{|c|c|c|c|}
\hline

process & incl. $\sigma$ & efficiency  &  $\sigma^{\textrm{rec}}$ \\
\hline

$t\bar t h$, single-lepton   & $111$ fb & $0.0485$ & $5.37$ fb \\
$t\bar t h$, di-lepton       & $17.7$ fb  &   $0.0359$  &   0.634 fb \\
\hline
$t\bar t $+jets, single-lepton&  $256$ pb & $0.463\times 10^{-3}$ & 119 fb  \\
$t\bar t$+jets, di-lepton    & $40.9$ pb  &  $0.168 \times 10^{-3}$ & 6.89 fb  \\
\hline

\end{tabular}
\end{center}
\caption{Total cross sections at the LHC 14 TeV and corresponding efficiency factors of the applied selection. }
\label{tab:xsec}
\end{table}

The event selection procedure is modeled after that 
adopted by the CMS collaboration for the measurement 
of the $t \bar t$ cross section in the 
di-lepton  channel~\cite{:2012qka}.
Single-lepton (di-lepton) events are required to have at least
one (one pair of opposite-charge) lepton(s).
Only isolated electrons or muons are lepton candidates
in our analysis. They are required to have 
a transverse momentum $p_T > 20$ GeV and a pseudo-rapidity
$|\eta | <2.4$. 
 
Jets are reconstructed via the anti-$k_T$ algorithm (with a cone radius $R=0.5$) as implemented in
{\sc FastJet}~\cite{Cacciari:2011ma} and applied on the calorimeter cells fired 
by the generated stable or quasi-stable particles. 
Jet candidates are required to have $p_T > 30$ GeV
and $|\eta|<2.5$, and  not to overlap with any selected
leptons. At least four $b$-jets are required. They are identified  with an efficiency $\epsilon_b= 0.7$ while
 mis-tag rates  $\epsilon_c= 0.2$ for charm quarks and  $\epsilon_j= 0.015$  for light partons are used. 
The cross sections for signal  and backgrounds together with the
final efficiencies of the adopted minimal selection procedure are collected in Table~\ref{tab:xsec}.

Only transfer functions for the jet energies are taken with a finite resolution,
which we parametrize through a double-Gaussian shape
function characterized by five independent parameters. 
For each of them, an energy dependence $c_1 + c_2 \sqrt{E}+ c_3 E$ is used. The 
constants $c_i$ are determined from
an independent $t \bar t$ sample where well 
separated jets (including light and $b$-jets) 
are matched to the corresponding partons. 
The typical resolution for the jet
energy is between 5 and 12 GeV, with tails parametrized by Gaussians of 
width as large as 30 GeV.

\begin{figure}[t]
\includegraphics[scale=0.22]{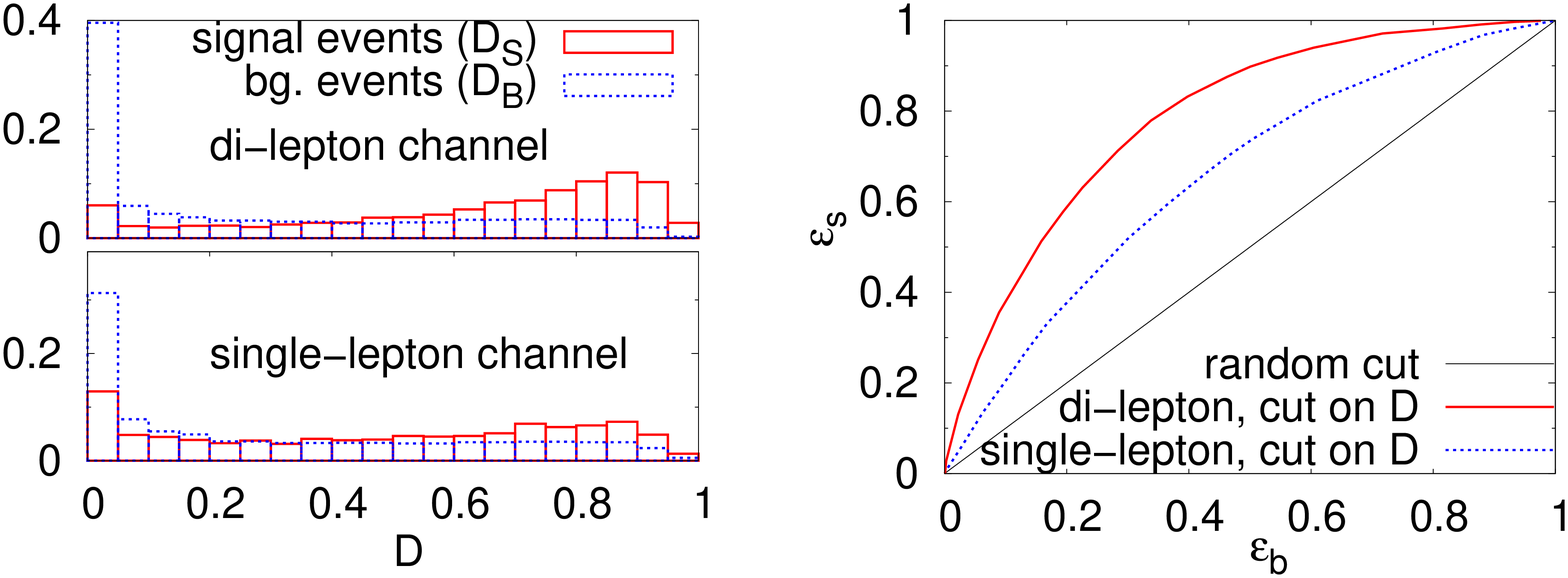}
\caption{Left: Normalized
distributions of events with respect to the MEM-based observable $D$ for the 
di-lepton (top) and single-lepton (bottom) channels. 
Right: Efficiency of selecting signal vs. background using a $D>D_{min}$ cut.
\label{discriminant}}
\end{figure}

\section{Results}

For a generic event $i$ with kinematics $\bs x_i$ the MEM-based observable $D_i$ is defined as follows:
\begin{equation}
D_i=\frac{P(\bs x_i|S)}{P(\bs x_i|S)+P(\bs x_i |B)}\,.
\end{equation}
Expected (normalized) distributions of signal and background events 
with respect to this observable are named $D_S$ and $D_B$, and are shown  
in Fig.~\ref{discriminant} (left). 
The plots show that for the same number 
of signal events  the MEM-based observable  delivers a higher 
discriminating power in the case of the di-lepton channel.
This is manifest in the right-hand plot of the same figure
where the $\epsilon_s$ versus $\epsilon_b$ efficiencies 
resulting from a cut on the observable $D>D_{min}$ are shown.
This may seem surprising at first sight, given that 
the di-lepton channel is characterized by two missing particles
in the final state, against only one in the single-lepton channel.
However, the di-lepton channel is much cleaner, with only $b$-jets required in the final
state, a lower probability of erroneously including extra QCD radiation  
and, eventually, a more manageable combinatorial background. 

In order to assess the significance that can be achieved at the 
LHC $\sqrt{s}=14$ TeV for a given luminosity $\mathcal{L}$, 
we consider a large number of pseudo-experiments, each with a number 
of events set to $N= \sigma^{\textrm{rec}}_{\textrm{bg}} \, \mathcal{L}$ 
(with $\sigma^{\textrm{rec}}_{\textrm{bg}}$ the reconstructed cross section, 
see Table~\ref{tab:xsec}, last column).
In the $B$-only hypothesis, 
the number of signal and background events are set to $s=0$ and $b=N$.
In the $S+B$ hypothesis, $s$ and $b$
are generated under the constraint $s+b=N$ according to the product
of Poisson distributions with mean values $Ns_0/(s_0+b_0)$ and $Nb_0/(s_0+b_0)$,
respectively. Here $s_0$ and $b_0$ are the expected number of reconstructed
events after rescaling the signal cross section by a parameter $\mu$, 
\textit{i.e.} $b_0= \sigma^{\textrm{rec}}_{\textrm{bg}} \, \mathcal{L}$ and
$s_0= \mu \, \sigma^{\textrm{rec}}_{\textrm{sig}} \,  \mathcal{L}$. 
For each event,  the corresponding 
$D_i$ value is generated according to
the probability law $D_S$ (in the case of a signal event)
or $D_B$ (in the case of a background event) shown in
Fig.~\ref{discriminant}.
This procedure is used to generate $10^4$ pseudo-experiments
under each hypothesis (B-only or S+B) at a given luminosity $\mathcal{L}$.

\begin{figure}[t]
\includegraphics[scale=0.7]{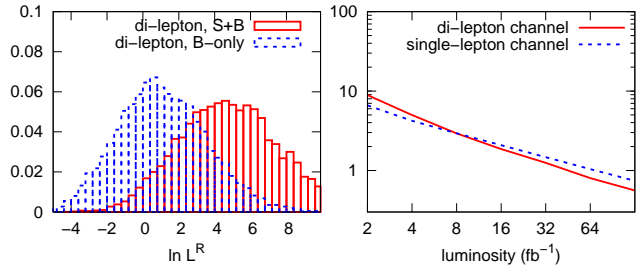}
\caption{
Left: Log likelihood profiles in the case of the di-lepton channel,
assuming a luminosity of 32 fb$^{-1}$ at 14 TeV and setting
$\mu=1$ (SM cross section). 
Right: Expected upper bound on the $t \bar t h$ cross section
(in units of SM cross section) at 95 \% C.L. 
\label{significance}}
\end{figure}

For each pseudo-experiment the likelihood ratio
$L^R$ is calculated as follows:
\begin{eqnarray}
 L^R &=&  \prod_{i}^N \frac{r_0P(\bs x_i |S )+(1-r_0)P(\bs x_i| B)}{P(\bs x_i| B)}  \label{Lshape} \nonumber \\
  &=& \prod_i^N   \frac{r_0  D_i +(1-r_0) (1-D_i)}{(1-D_i)}   \,,
\end{eqnarray}
with $r_0=s_0/(s_0+b_0)$.
The resulting $B$-only and $S+B$ distributions of pseudo-experiments with respect to 
$\ln \left( L^R  \right)$ are shown in Fig.~\ref{significance} (left)
in the case of the di-lepton channel, with $\mathcal{L}=$32 fb$^{-1}$ and 
$\mu=1$. The two distributions are shifted towards positive values of 
$\ln\left( L^R \right)$, which indicates that the MEM weights 
do not exactly describe the phase-space distributions 
of background and signal events. This bias originates 
from the approximations inherent to the calculation of the 
weights, e.g.,   the assumed parametrization of the transfer function
and the effective treatment of beyond-leading-order  QCD radiations. 

By  smearing the value of $b_0$
according to a log-normal distribution (mean=$b_0$, std=$0.2b_0$)
before generating $s$ and $b$ in each pseudo-experiment, 
we also verified that 
 systematic uncertainties on the background normalization 
have a negligible impact on the distributions of pseudo-experiments 
with respect to $\ln\left( L^R \right)$. On the other hand, 
already a $20$\% uncertainty on $b_0$ hampers a counting analysis 
based on the number of events to be available at LHC, unless $s/b \gg 0.2$.

We repeat this exercise with different values of $\mu$
until the median of the $B$-only distribution cuts 
5\% of the left-hand tail of the $S+B$ distribution. 
Such a value of $\mu$ provides us with the
estimate $\mu \times \sigma(t\bar t h)$ of the expected 
upper bound on the signal cross section at 95 \% C.L.
in the absence of signal.
Fig.~\ref{significance} (right) shows our estimate of the parameter $\mu$ as a function
of the luminosity $\mathcal{L}$, 
separately for the di-lepton and single-lepton channels.
We observe that the sensitivity achieved in the di-lepton 
channel is slightly better than the one in the single-lepton channel at large luminosities.  

\section{Conclusions}
The matrix element reweighting method is a powerful technique to enhance
the sensitivity of searches and measurements at colliders.
In this work we have applied it to the observation of Higgs production in association with $t\bar t$ pair,
in particular to  final state  signatures involving the decay of the Higgs to bottom quarks.  
First, we have verified that the general algorithm implemented in {\sc MadWeight} provides the possibility of automatically, reliably and quickly calculating weights for  final states as complex as those featured in $t \bar t h$. This technical result, by itself, is an important one as it opens the door to applications of the MEM to a much wider set of processes and analyses than what has been done so far.
Second, we have applied the method to $t\bar t h$ with both one- and two-lepton final states. 
We have found that the di-lepton final state,  though penalized by a smaller number of expected events and
possibly more difficult to reconstruct due to the presence of two neutrinos in the final state,  becomes competitive with the 
single-lepton channel already after a moderate integrated luminosity. 
We reckon that  this result, while based on MC simulations, is rather robust and  encourage more refined
investigations.  
For instance, relaxing the number of requested $b$-tags to three in the di-lepton final state would bring a
significant increase in the statistics, yet not of the combinatorial background,  leading to a further relative
gain with respect to the single-lepton final state. Analogously, the inclusion of pile-up will impact more the
signature with the largest number of jets in the final state.
Only fully-fledged experimental analyses can assess
the final gains, and correctly include the systematic uncertainties  that have been neglected here. 

In conclusion the search for SM $t\bar t h$ and in particular the di-lepton final state is a perfect illustration 
of the power of the matrix element method in providing additional leverage in difficult analyses.  Further investigations
concerning the possibility of using the matrix element method in $t\bar t h$ to access more detailed information on the structure of the 
couplings of the new boson or in other very challenging production channels, such at $thj$~\cite{Farina:2012xp},  are foreseen.

\section{Acknowledgements}
We are grateful to J. D' Hondt, K. Cranmer, R. Frederix, and  to many CP3 members 
for very useful information exchange and discussions.  P.A. is supported by a 
Marie Curie Intra-European Fellowship (PIEF-GA-2011-299999 PROBE4TeVSCALE).
 P.d.A. is supported in part by the Belgian Federal Science Policy
Office through the Interuniversity Attraction Pole P7/37, in part by the
``FWO-Vlaanderen" through the project G.0114.10N, and in part by the 
Concerted Research action "Supersymmetric Models and their 
Signatures at the Large Hadron Collider" of the Vrije Universiteit Brussel (VUB)".
O.M. is fellow of the Belgian American Education Foundation.

\bibliography{tth}

\begin{thebibliography}{31}
\expandafter\ifx\csname natexlab\endcsname\relax\def\natexlab#1{#1}\fi
\expandafter\ifx\csname bibnamefont\endcsname\relax
  \def\bibnamefont#1{#1}\fi
\expandafter\ifx\csname bibfnamefont\endcsname\relax
  \def\bibfnamefont#1{#1}\fi
\expandafter\ifx\csname citenamefont\endcsname\relax
  \def\citenamefont#1{#1}\fi
\expandafter\ifx\csname url\endcsname\relax
  \def\url#1{\texttt{#1}}\fi
\expandafter\ifx\csname urlprefix\endcsname\relax\def\urlprefix{URL }\fi
\providecommand{\bibinfo}[2]{#2}
\providecommand{\eprint}[2][]{\url{#2}}

\bibitem[{cms(2013)}]{cms2013}
\bibinfo{journal}{CMS-HIG-13-003. CMS-HIG-13-004. CMS-HIG-13-006.
  CMS-HIG-13-009.}  (\bibinfo{year}{2013}).

\bibitem[{atl(2013)}]{atlas2013}
\bibinfo{journal}{ATLAS-CONF-2013-009. ATLAS-CONF-2013-010.
  ATLAS-CONF-2013-012. ATLAS- CONF-2013-013.}  (\bibinfo{year}{2013}).

\bibitem[{\citenamefont{Degrande et~al.}(2012)\citenamefont{Degrande, Gerard,
  Grojean, Maltoni, and Servant}}]{Degrande:2012gr}
\bibinfo{author}{\bibfnamefont{C.}~\bibnamefont{Degrande}},
  \bibinfo{author}{\bibfnamefont{J.}~\bibnamefont{Gerard}},
  \bibinfo{author}{\bibfnamefont{C.}~\bibnamefont{Grojean}},
  \bibinfo{author}{\bibfnamefont{F.}~\bibnamefont{Maltoni}}, \bibnamefont{and}
  \bibinfo{author}{\bibfnamefont{G.}~\bibnamefont{Servant}},
  \bibinfo{journal}{JHEP} \textbf{\bibinfo{volume}{1207}}, \bibinfo{pages}{036}
  (\bibinfo{year}{2012}), \eprint{1205.1065}.

\bibitem[{\citenamefont{Frederix et~al.}(2011)\citenamefont{Frederix, Frixione,
  Hirschi, Maltoni, Pittau, and Torrielli}}]{Frederix:2011zi}
\bibinfo{author}{\bibfnamefont{R.}~\bibnamefont{Frederix}},
  \bibinfo{author}{\bibfnamefont{S.}~\bibnamefont{Frixione}},
  \bibinfo{author}{\bibfnamefont{V.}~\bibnamefont{Hirschi}},
  \bibinfo{author}{\bibfnamefont{F.}~\bibnamefont{Maltoni}},
  \bibinfo{author}{\bibfnamefont{R.}~\bibnamefont{Pittau}}, \bibnamefont{and}
  \bibinfo{author}{\bibfnamefont{P.}~\bibnamefont{Torrielli}},
  \bibinfo{journal}{Phys.Lett.} \textbf{\bibinfo{volume}{B701}},
  \bibinfo{pages}{427} (\bibinfo{year}{2011}), \eprint{1104.5613}.

\bibitem[{\citenamefont{Beenakker et~al.}(2001)\citenamefont{Beenakker,
  Dittmaier, Kramer, Plumper, Spira et~al.}}]{Beenakker:2001rj}
\bibinfo{author}{\bibfnamefont{W.}~\bibnamefont{Beenakker}},
  \bibinfo{author}{\bibfnamefont{S.}~\bibnamefont{Dittmaier}},
  \bibinfo{author}{\bibfnamefont{M.}~\bibnamefont{Kramer}},
  \bibinfo{author}{\bibfnamefont{B.}~\bibnamefont{Plumper}},
  \bibinfo{author}{\bibfnamefont{M.}~\bibnamefont{Spira}},
  \bibnamefont{et~al.}, \bibinfo{journal}{Phys.Rev.Lett.}
  \textbf{\bibinfo{volume}{87}}, \bibinfo{pages}{201805}
  (\bibinfo{year}{2001}), \eprint{hep-ph/0107081}.

\bibitem[{\citenamefont{Dawson et~al.}(2003)\citenamefont{Dawson, Orr, Reina,
  and Wackeroth}}]{Dawson:2002tg}
\bibinfo{author}{\bibfnamefont{S.}~\bibnamefont{Dawson}},
  \bibinfo{author}{\bibfnamefont{L.}~\bibnamefont{Orr}},
  \bibinfo{author}{\bibfnamefont{L.}~\bibnamefont{Reina}}, \bibnamefont{and}
  \bibinfo{author}{\bibfnamefont{D.}~\bibnamefont{Wackeroth}},
  \bibinfo{journal}{Phys.Rev.} \textbf{\bibinfo{volume}{D67}},
  \bibinfo{pages}{071503} (\bibinfo{year}{2003}), \eprint{hep-ph/0211438}.

\bibitem[{\citenamefont{Garzelli et~al.}(2011)\citenamefont{Garzelli, Kardos,
  Papadopoulos, and Trocsanyi}}]{Garzelli:2011vp}
\bibinfo{author}{\bibfnamefont{M.}~\bibnamefont{Garzelli}},
  \bibinfo{author}{\bibfnamefont{A.}~\bibnamefont{Kardos}},
  \bibinfo{author}{\bibfnamefont{C.}~\bibnamefont{Papadopoulos}},
  \bibnamefont{and}
  \bibinfo{author}{\bibfnamefont{Z.}~\bibnamefont{Trocsanyi}},
  \bibinfo{journal}{Europhys.Lett.} \textbf{\bibinfo{volume}{96}},
  \bibinfo{pages}{11001} (\bibinfo{year}{2011}), \eprint{1108.0387}.

\bibitem[{\citenamefont{Maltoni et~al.}(2002)\citenamefont{Maltoni, Rainwater,
  and Willenbrock}}]{Maltoni:2002jr}
\bibinfo{author}{\bibfnamefont{F.}~\bibnamefont{Maltoni}},
  \bibinfo{author}{\bibfnamefont{D.~L.} \bibnamefont{Rainwater}},
  \bibnamefont{and}
  \bibinfo{author}{\bibfnamefont{S.}~\bibnamefont{Willenbrock}},
  \bibinfo{journal}{Phys.Rev.} \textbf{\bibinfo{volume}{D66}},
  \bibinfo{pages}{034022} (\bibinfo{year}{2002}), \eprint{hep-ph/0202205}.

\bibitem[{\citenamefont{Belyaev and Reina}(2002)}]{Belyaev:2002ua}
\bibinfo{author}{\bibfnamefont{A.}~\bibnamefont{Belyaev}} \bibnamefont{and}
  \bibinfo{author}{\bibfnamefont{L.}~\bibnamefont{Reina}},
  \bibinfo{journal}{JHEP} \textbf{\bibinfo{volume}{0208}}, \bibinfo{pages}{041}
  (\bibinfo{year}{2002}), \eprint{hep-ph/0205270}.

\bibitem[{\citenamefont{Buttar et~al.}(2006)\citenamefont{Buttar, Dittmaier,
  Drollinger, Frixione, Nikitenko et~al.}}]{Buttar:2006zd}
\bibinfo{author}{\bibfnamefont{C.}~\bibnamefont{Buttar}},
  \bibinfo{author}{\bibfnamefont{S.}~\bibnamefont{Dittmaier}},
  \bibinfo{author}{\bibfnamefont{V.}~\bibnamefont{Drollinger}},
  \bibinfo{author}{\bibfnamefont{S.}~\bibnamefont{Frixione}},
  \bibinfo{author}{\bibfnamefont{A.}~\bibnamefont{Nikitenko}},
  \bibnamefont{et~al.} (\bibinfo{year}{2006}), \eprint{hep-ph/0604120}.

\bibitem[{\citenamefont{Plehn et~al.}(2010)\citenamefont{Plehn, Salam, and
  Spannowsky}}]{Plehn:2009rk}
\bibinfo{author}{\bibfnamefont{T.}~\bibnamefont{Plehn}},
  \bibinfo{author}{\bibfnamefont{G.~P.} \bibnamefont{Salam}}, \bibnamefont{and}
  \bibinfo{author}{\bibfnamefont{M.}~\bibnamefont{Spannowsky}},
  \bibinfo{journal}{Phys.Rev.Lett.} \textbf{\bibinfo{volume}{104}},
  \bibinfo{pages}{111801} (\bibinfo{year}{2010}), \eprint{0910.5472}.

\bibitem[{\citenamefont{Kondo}(1988)}]{Kondo:1988yd}
\bibinfo{author}{\bibfnamefont{K.}~\bibnamefont{Kondo}},
  \bibinfo{journal}{J.Phys.Soc.Jap.} \textbf{\bibinfo{volume}{57}},
  \bibinfo{pages}{4126} (\bibinfo{year}{1988}).

\bibitem[{\citenamefont{Abazov et~al.}(2004)}]{Abazov:2004cs}
\bibinfo{author}{\bibfnamefont{V.}~\bibnamefont{Abazov}} \bibnamefont{et~al.}
  (\bibinfo{collaboration}{D0 Collaboration}), \bibinfo{journal}{Nature}
  \textbf{\bibinfo{volume}{429}}, \bibinfo{pages}{638} (\bibinfo{year}{2004}),
  \eprint{hep-ex/0406031}.

\bibitem[{\citenamefont{Abulencia et~al.}(2007)}]{Abulencia:2006ry}
\bibinfo{author}{\bibfnamefont{A.}~\bibnamefont{Abulencia}}
  \bibnamefont{et~al.} (\bibinfo{collaboration}{CDF Collaboration}),
  \bibinfo{journal}{Phys.Rev.} \textbf{\bibinfo{volume}{D75}},
  \bibinfo{pages}{031105} (\bibinfo{year}{2007}), \eprint{hep-ex/0612060}.

\bibitem[{\citenamefont{Aaltonen et~al.}(2009)}]{Aaltonen:2009jj}
\bibinfo{author}{\bibfnamefont{T.}~\bibnamefont{Aaltonen}} \bibnamefont{et~al.}
  (\bibinfo{collaboration}{CDF Collaboration}),
  \bibinfo{journal}{Phys.Rev.Lett.} \textbf{\bibinfo{volume}{103}},
  \bibinfo{pages}{092002} (\bibinfo{year}{2009}), \eprint{0903.0885}.

\bibitem[{\citenamefont{Abazov et~al.}(2009)}]{Abazov:2009ii}
\bibinfo{author}{\bibfnamefont{V.}~\bibnamefont{Abazov}} \bibnamefont{et~al.}
  (\bibinfo{collaboration}{D0 Collaboration}),
  \bibinfo{journal}{Phys.Rev.Lett.} \textbf{\bibinfo{volume}{103}},
  \bibinfo{pages}{092001} (\bibinfo{year}{2009}), \eprint{0903.0850}.

\bibitem[{\citenamefont{Aad et~al.}(2012)}]{Aad:2012tfa}
\bibinfo{author}{\bibfnamefont{G.}~\bibnamefont{Aad}} \bibnamefont{et~al.}
  (\bibinfo{collaboration}{ATLAS Collaboration}), \bibinfo{journal}{Phys.Lett.}
  \textbf{\bibinfo{volume}{B716}}, \bibinfo{pages}{1} (\bibinfo{year}{2012}),
  \eprint{1207.7214}.

\bibitem[{\citenamefont{Chatrchyan
  et~al.}(2012{\natexlab{a}})}]{Chatrchyan:2012ufa}
\bibinfo{author}{\bibfnamefont{S.}~\bibnamefont{Chatrchyan}}
  \bibnamefont{et~al.} (\bibinfo{collaboration}{CMS Collaboration}),
  \bibinfo{journal}{Phys.Lett.} \textbf{\bibinfo{volume}{B716}},
  \bibinfo{pages}{30} (\bibinfo{year}{2012}{\natexlab{a}}), \eprint{1207.7235}.

\bibitem[{\citenamefont{Campbell et~al.}(2012)\citenamefont{Campbell, Giele,
  and Williams}}]{Campbell:2012cz}
\bibinfo{author}{\bibfnamefont{J.~M.} \bibnamefont{Campbell}},
  \bibinfo{author}{\bibfnamefont{W.~T.} \bibnamefont{Giele}}, \bibnamefont{and}
  \bibinfo{author}{\bibfnamefont{C.}~\bibnamefont{Williams}},
  \bibinfo{journal}{JHEP} \textbf{\bibinfo{volume}{1211}}, \bibinfo{pages}{043}
  (\bibinfo{year}{2012}), \eprint{1204.4424}.

\bibitem[{\citenamefont{Campbell et~al.}(2013)\citenamefont{Campbell, Ellis,
  Giele, and Williams}}]{Campbell:2013hz}
\bibinfo{author}{\bibfnamefont{J.~M.} \bibnamefont{Campbell}},
  \bibinfo{author}{\bibfnamefont{R.~K.} \bibnamefont{Ellis}},
  \bibinfo{author}{\bibfnamefont{W.~T.} \bibnamefont{Giele}}, \bibnamefont{and}
  \bibinfo{author}{\bibfnamefont{C.}~\bibnamefont{Williams}}
  (\bibinfo{year}{2013}), \eprint{1301.7086}.

\bibitem[{\citenamefont{Artoisenet et~al.}(2010)\citenamefont{Artoisenet,
  Lemaitre, Maltoni, and Mattelaer}}]{Artoisenet:2010cn}
\bibinfo{author}{\bibfnamefont{P.}~\bibnamefont{Artoisenet}},
  \bibinfo{author}{\bibfnamefont{V.}~\bibnamefont{Lemaitre}},
  \bibinfo{author}{\bibfnamefont{F.}~\bibnamefont{Maltoni}}, \bibnamefont{and}
  \bibinfo{author}{\bibfnamefont{O.}~\bibnamefont{Mattelaer}},
  \bibinfo{journal}{JHEP} \textbf{\bibinfo{volume}{1012}}, \bibinfo{pages}{068}
  (\bibinfo{year}{2010}), \eprint{1007.3300}.

\bibitem[{\citenamefont{Alwall et~al.}(2011{\natexlab{a}})\citenamefont{Alwall,
  Freitas, and Mattelaer}}]{Alwall:2010cq}
\bibinfo{author}{\bibfnamefont{J.}~\bibnamefont{Alwall}},
  \bibinfo{author}{\bibfnamefont{A.}~\bibnamefont{Freitas}}, \bibnamefont{and}
  \bibinfo{author}{\bibfnamefont{O.}~\bibnamefont{Mattelaer}},
  \bibinfo{journal}{Phys.Rev.} \textbf{\bibinfo{volume}{D83}},
  \bibinfo{pages}{074010} (\bibinfo{year}{2011}{\natexlab{a}}),
  \eprint{1010.2263}.

\bibitem[{\citenamefont{Alwall et~al.}(2011{\natexlab{b}})\citenamefont{Alwall,
  Herquet, Maltoni, Mattelaer, and Stelzer}}]{Alwall:2011uj}
\bibinfo{author}{\bibfnamefont{J.}~\bibnamefont{Alwall}},
  \bibinfo{author}{\bibfnamefont{M.}~\bibnamefont{Herquet}},
  \bibinfo{author}{\bibfnamefont{F.}~\bibnamefont{Maltoni}},
  \bibinfo{author}{\bibfnamefont{O.}~\bibnamefont{Mattelaer}},
  \bibnamefont{and} \bibinfo{author}{\bibfnamefont{T.}~\bibnamefont{Stelzer}},
  \bibinfo{journal}{JHEP} \textbf{\bibinfo{volume}{1106}}, \bibinfo{pages}{128}
  (\bibinfo{year}{2011}{\natexlab{b}}), \eprint{1106.0522}.

\bibitem[{\citenamefont{Sjostrand et~al.}(2006)\citenamefont{Sjostrand, Mrenna,
  and Skands}}]{Sjostrand:2006za}
\bibinfo{author}{\bibfnamefont{T.}~\bibnamefont{Sjostrand}},
  \bibinfo{author}{\bibfnamefont{S.}~\bibnamefont{Mrenna}}, \bibnamefont{and}
  \bibinfo{author}{\bibfnamefont{P.~Z.} \bibnamefont{Skands}},
  \bibinfo{journal}{JHEP} \textbf{\bibinfo{volume}{0605}}, \bibinfo{pages}{026}
  (\bibinfo{year}{2006}), \eprint{hep-ph/0603175}.

\bibitem[{\citenamefont{Alwall et~al.}(2009)\citenamefont{Alwall, de~Visscher,
  and Maltoni}}]{Alwall:2008qv}
\bibinfo{author}{\bibfnamefont{J.}~\bibnamefont{Alwall}},
  \bibinfo{author}{\bibfnamefont{S.}~\bibnamefont{de~Visscher}},
  \bibnamefont{and} \bibinfo{author}{\bibfnamefont{F.}~\bibnamefont{Maltoni}},
  \bibinfo{journal}{JHEP} \textbf{\bibinfo{volume}{0902}}, \bibinfo{pages}{017}
  (\bibinfo{year}{2009}), \eprint{0810.5350}.

\bibitem[{\citenamefont{Ovyn et~al.}(2009)\citenamefont{Ovyn, Rouby, and
  Lemaitre}}]{Ovyn:2009tx}
\bibinfo{author}{\bibfnamefont{S.}~\bibnamefont{Ovyn}},
  \bibinfo{author}{\bibfnamefont{X.}~\bibnamefont{Rouby}}, \bibnamefont{and}
  \bibinfo{author}{\bibfnamefont{V.}~\bibnamefont{Lemaitre}}
  (\bibinfo{year}{2009}), \eprint{0903.2225}.

\bibitem[{\citenamefont{Cacciari
  et~al.}(2012{\natexlab{a}})\citenamefont{Cacciari, Czakon, Mangano, Mitov,
  and Nason}}]{Cacciari:2011hy}
\bibinfo{author}{\bibfnamefont{M.}~\bibnamefont{Cacciari}},
  \bibinfo{author}{\bibfnamefont{M.}~\bibnamefont{Czakon}},
  \bibinfo{author}{\bibfnamefont{M.}~\bibnamefont{Mangano}},
  \bibinfo{author}{\bibfnamefont{A.}~\bibnamefont{Mitov}}, \bibnamefont{and}
  \bibinfo{author}{\bibfnamefont{P.}~\bibnamefont{Nason}},
  \bibinfo{journal}{Phys.Lett.} \textbf{\bibinfo{volume}{B710}},
  \bibinfo{pages}{612} (\bibinfo{year}{2012}{\natexlab{a}}),
  \eprint{1111.5869}.

\bibitem[{\citenamefont{Artoisenet et~al.}(2012)\citenamefont{Artoisenet,
  Frederix, Mattelaer, and Rietkerk}}]{Artoisenet:2012st}
\bibinfo{author}{\bibfnamefont{P.}~\bibnamefont{Artoisenet}},
  \bibinfo{author}{\bibfnamefont{R.}~\bibnamefont{Frederix}},
  \bibinfo{author}{\bibfnamefont{O.}~\bibnamefont{Mattelaer}},
  \bibnamefont{and} \bibinfo{author}{\bibfnamefont{R.}~\bibnamefont{Rietkerk}}
  (\bibinfo{year}{2012}), \eprint{1212.3460}.

\bibitem[{\citenamefont{Chatrchyan et~al.}(2012{\natexlab{b}})}]{:2012qka}
\bibinfo{author}{\bibfnamefont{S.}~\bibnamefont{Chatrchyan}}
  \bibnamefont{et~al.} (\bibinfo{collaboration}{CMS Collaboration})
  (\bibinfo{year}{2012}{\natexlab{b}}), \eprint{1211.2220}.

\bibitem[{\citenamefont{Cacciari
  et~al.}(2012{\natexlab{b}})\citenamefont{Cacciari, Salam, and
  Soyez}}]{Cacciari:2011ma}
\bibinfo{author}{\bibfnamefont{M.}~\bibnamefont{Cacciari}},
  \bibinfo{author}{\bibfnamefont{G.~P.} \bibnamefont{Salam}}, \bibnamefont{and}
  \bibinfo{author}{\bibfnamefont{G.}~\bibnamefont{Soyez}},
  \bibinfo{journal}{Eur.Phys.J.} \textbf{\bibinfo{volume}{C72}},
  \bibinfo{pages}{1896} (\bibinfo{year}{2012}{\natexlab{b}}),
  \eprint{1111.6097}.

\bibitem[{\citenamefont{Farina et~al.}(2012)\citenamefont{Farina, Grojean,
  Maltoni, Salvioni, and Thamm}}]{Farina:2012xp}
\bibinfo{author}{\bibfnamefont{M.}~\bibnamefont{Farina}},
  \bibinfo{author}{\bibfnamefont{C.}~\bibnamefont{Grojean}},
  \bibinfo{author}{\bibfnamefont{F.}~\bibnamefont{Maltoni}},
  \bibinfo{author}{\bibfnamefont{E.}~\bibnamefont{Salvioni}}, \bibnamefont{and}
  \bibinfo{author}{\bibfnamefont{A.}~\bibnamefont{Thamm}}
  (\bibinfo{year}{2012}), \eprint{1211.3736}.

\end{thebibliography}

\end{document}